\documentclass[aps,pra,superscriptaddress,12pt,tightenlines,nofootinbib]{revtex4}
\usepackage{amsmath,amsthm,graphicx,amssymb,verbatim,listings}
\usepackage{wasysym}
\usepackage[T1]{fontenc}     
\usepackage{lmodern}         
\usepackage[ pdftex, plainpages = false, pdfpagelabels,
                 pdfpagelayout = useoutlines,
                 bookmarks,
                 bookmarksopen = true,
                 bookmarksnumbered = true,
                 breaklinks = true,
                 linktocpage=all,
                 pagebackref=false,
                 colorlinks = true,
                 linkcolor = BrickRed,
                 urlcolor  = blue,
                 citecolor = BrickRed,
                 anchorcolor = green,
                 hyperindex = true,
                 hyperfigures
                 ]{hyperref}
\usepackage[usenames, dvipsnames]{xcolor}
\usepackage{xifthen} 

\newcommand{\ket}[1]{{\ensuremath{\left| #1 \right\rangle}}}
\newcommand{\bra}[1]{{\ensuremath{\left\langle #1 \right|}}}

\newcommand{\arxiv}[2][]{\ifthenelse{\isempty{#1}}{\href{http://arxiv.org/abs/#2}{{\tt arXiv:\allowbreak{}#2}}} {\href{http://arxiv.org/abs/#2}{{\tt arXiv:\allowbreak{}#2 [#1]}}}}

\newcommand{\booktitle}{\textsl}
\newcommand{\hrefdoi}[2]{\href{https://dx.doi.org/#1}{#2}}

\newcommand{\cX}{\mathcal{X}}

\begin{document}

\title{On QBism and Assumption (Q)}

\author{Blake C.\ Stacey}
\affiliation{\href{http://www.physics.umb.edu/Research/QBism/}{QBism Research Group}, University of Massachusetts Boston}

\date{\today}
\begin{abstract}
I correct two misapprehensions, one historical and one conceptual, in
the recent literature on extensions of the Wigner's Friend
thought-experiment. Perhaps fittingly, both concern the accurate
description of some quantum physicists' beliefs by others.
\end{abstract}
\maketitle

This note is a response to Sudbery's recent
commentary~\cite{Sudbery:2019} on an article by Frauchiger and
Renner~\cite{Frauchiger:2018}, who propose an elaboration upon a
thought-experiment named in honor of Wigner~\cite{Wigner:1961}. With
that intimidatingly scholastic sentence out of the way, we will first
address a bibliographical point and then proceed to more conceptual
matters.

Sudbery's paper uses a 2002 article by Caves, Fuchs and
Schack~\cite{CFS:2002} to define the interpretation of quantum
mechanics known as QBism. This is an incorrect attribution. That paper
contains much that is not QBist, and all three authors will say
so~\cite{3}. QBism had not yet clawed itself into being in 2002, nor
even by 2007~\cite{CFS:2007}. QBism, as a term and an interpretation,
did not hit the arXiv until 2009~\cite[\S 8]{Fuchs:2009}. Moreover,
Caves does not call himself a QBist and disagrees with some turns that
the other two authors made in the process of developing the QBist
interpretation. (One might call the Caves--Fuchs--Schack
collaborations ``Quantum Bayesian'' papers, that term being older and
more general, encompassing some writings of Bub and
Pitowsky~\cite{Bub:2009} and potentially even old Usenet posts of
Baez~\cite{Baez:2003}, for example.)  Canonical presentations of
QBism, in roughly increasing order of technicality, include von
Baeyer's book~\cite{VonBaeyer:2016}; Fuchs, Mermin and Schack's
article~\cite{Fuchs:2013c}; and Fuchs and Stacey's long
review~\cite{Fuchs:2019}. What these sources say about Wigner's Friend
applies to Frauchiger--Renner and all the other recent variations
thereof.

Nurgalieva and del Rio's analysis of
Frauchiger--Renner~\cite{Nurgalieva:2018} also cites the 2002 paper of
Caves, Fuchs and Schack as QBist, but provides correct references in
addition.

With that addressed, we move on.  Sudbery gives the following
paraphrase of Frauchiger and Renner's Assumption (Q):
\begin{quotation}
  \noindent If an agent $A$ is certain that a system is in an
  eigenstate of an observable $X$ with eigenvalue $\xi$ and time
  $t_0$, and a measurement of $X$ is completed at time $t > t_0$, then
  $A$ is certain that the result $x$ of the measurement is $x = \xi$
  at time $t$.
\end{quotation}
Sudbery asserts that QBism accepts this assumption (see his Table
1). This is incorrect; in fact, QBism finds it ill-posed in multiple
ways.  First, a QBist would never say that a system is ``\emph{in}'' a
quantum state, eigen- or otherwise.  Any quantum state is an agent's
encoding of their own expectations regarding a physical system.  The
system is not \emph{in} a quantum state any more than the Earth's
atmosphere is \emph{in} a weather forecast. Therefore, a good QBist
agent would not be ``certain that a system is in an eigenstate'' of
$X$; the agent's choice of a pure-state ascription is itself the
expression of their certainty. This point also affects the evaluation
of how QBism responds to other statements, such as Sudbery's
Assumption (P).

Second, and more subtly, Assumption (Q) conflates statements applying
to different times. Let us say that at time $t_0$, Alice ascribes the
quantum state $\ket{\psi}$ to a physical system of interest.  If Alice
lives up to the consistency standard known as quantum theory, then she
must consequently be maximally confident regarding the outcome of a
specific later measurement.  At time $t_0$, she fully expects that the
von Neumann measurement $X$ completed at time $t > t_0$ will have
outcome $\xi$.  This is a statement about her gambling commitments at
time $t_0$, not what she might believe when time $t$ finally
arrives. There is a crucial difference between ``Alice is certain at
time $t$ that $X$ will yield $\xi$'' and ``Alice is certain at time $t_0$
\emph{that she will be certain at a later time} $t$ that $X$ will yield
$\xi$''. For additional discussion deliniating the qualitatively
different things too often clumped together as ``time evolution'',
see~\cite{Fuchs:2012, Stacey:2015, DeBrota:2018}.

Here is Frauchiger and Renner's version of Assumption (Q):
\begin{quotation}
  \noindent Suppose that agent A has established that
  \begin{quote}
  \emph{Statement} A\textsuperscript{(i)}: ``System S is in state
  $\ket{\psi_S}$ at time $t_0$.''
  \end{quote}
  Suppose furthermore that agent A knows that
  \begin{quote}
  \emph{Statement} A\textsuperscript{(ii)}: ``The value $x$ is obtained by
  a measurement of S w.r.t. the family $\{\pi_x^{t_0}\}_{x\in \cX}$ of
  Heisenberg operators relative to time $t_0$, which is completed at time
  $t$.''
  \end{quote}
  If $\bra{\psi} \pi_\xi^{t_0} \ket{\psi} = 1$ for some $\xi \in \cX$
  then agent A can conclude that
  \begin{quote}
  \emph{Statement} A\textsuperscript{(iii)}: ``I am certain that $x =
  \xi$ at time $t$.''
  \end{quote}
\end{quotation}
This version is more notationally elaborate, but as far as QBism is
concerned, it shares the same problems as Sudbery's more succinct
phrasing.  There is an additional difficulty with Frauchiger and
Renner's discussion:
\begin{quotation}
  \noindent Assumption (Q) corresponds to the quantum-mechanical Born
  rule. Since the assumption is concerned with the special case of
  probability-1 predictions only, it is largely independent of
  interpretational questions, such as the meaning of probabilities in
  general.
\end{quotation}
QBism maintains that predictions made with $p = 1$ are still
\emph{probabilistic} predictions, just as those with $p = 0.99$, or $p
= 0.999$. This is an essential ingredient in the QBist take on the EPR
``paradox'' and Bell inequalities~\cite{Fuchs:2013c,
  Fuchs:2019}. Statement A\textsuperscript{(iii)} in Assumption (Q)
appears to play the role of the EPR criterion of reality, namely
attempting to infer pre-existing physical reality from $p = 1$
predictions, a move that QBism finds inadmissable. Whether or not one
accepts this position, the fact is that it exists in the spectrum of
foundational debate, and taking it on board affects the validity of
conclusions one might attempt to draw. Thus, restricting one's
argument to probability-1 predictions does not make one's desired
conclusion ``largely independent'' of the interpretation of
probability.  The question of how to interpret probability theory is
too fundamental to be sidestepped in this manner. Trying to do so is
like saying, ``We discuss only the colors black and white, so our
logic is independent of questions about visual perception.''

Consequently, the row about QBism in Frauchiger and Renner's Table 4
is in error.

Further examination and exegesis are possible~\cite{DeBrota:2018}. For
example, the question of interpreting probability-1 statements is a
good jumping-off point to understand the divergence between the
Caves--Fuchs--Schack collaborations and QBism
proper~\cite{Fuchs:2014}. However, this comment is already long
enough.

\bigskip

I thank Frauchiger and Renner for bringing the old riddle of Wigner's
Friend into the limelight, and I also extend my gratitude to del Rio,
Nurgalieva and Sudbery for taking on the typically thankless task of
articulating assumptions otherwise left unstated. Moreover, I have had
helpful conversations and correspondence with Gabriela Barreto Lemos,
Carlton Caves, John B.\ DeBrota, Christopher Fuchs, Jacques Pienaar
and R\"udiger Schack.

This research was supported by the John Templeton Foundation. The
opinions expressed in this publication are those of the author and do
not necessarily reflect the views of the John Templeton Foundation.

\end{document}